\documentclass[useAMS,usenatbib]{mn2e}
\usepackage{psfrag,graphicx}
\usepackage{color}
\usepackage{txfonts}
\usepackage{breqn}
\usepackage[T1]{fontenc}
\usepackage{ae,aecompl}

\title[Black hole formation in the early universe]
{Black hole formation in the context of dissipative dark matter}
 \author[Latif et al. ]
{M. A. Latif\thanks{Corresponding author: latifne@gmail.com}$^1$,
A. Lupi$^2$,
D. R. G. Schleicher$^3$,
G. D'Amico$^{4,6}$,
P. Panci$^{4,5}$,
S. Bovino$^3$ \\
$^1$Physics Department, College of Science, United Arab Emirates University, PO Box 15551, Al-Ain, UAE\\
$^2$Scuola Normale Superiore, Piazza dei Cavalieri, 7, 56126, PISA, Italy \\
$^3$Departamento de Astronomía, Facultad Ciencias Físicas y Matemáticas, Universidad de Concepción, \\
       Av. Esteban Iturra s/n Barrio Universitario, Casilla 160-C, Chile  \\
$^4$CERN Theoretical Physics Department, Case C01600, CH-1211 Geneve, Switzerland \\
$^5$Laboratori Nazionali del Gran Sasso, Via G. Acitelli, 22, I-67100 Assergi (AQ), Italy \\
$^6$Stanford Institute for Theoretical Physics, Stanford University, Stanford, CA 94306, USA
}
\date{}

\def\LaTeX{L\kern-.36em\raise.3ex\hbox{a}\kern-.15em
      T\kern-.1667em\lower.7ex\hbox{E}\kern-.125emX}

\begin{document}

\bibliographystyle{mn2e}

\label{firstpage}

\maketitle
\def\na{NewA}%
\def\aj{AJ}%
\def\araa{ARA\&A}%
\def\apj{ApJ}%
\def\apjl{ApJ}%
\def\jcap{JCAP}

\def\apjs{ApJS}%
\def\ao{Appl.~Opt.}%
\def\apss{Ap\&SS}%
\def\aap{A\&A}%
\def\aapr{A\&A~Rev.}%
\def\aaps{A\&AS}%
\def\azh{AZh}%
\def\baas{BAAS}%
\def\jrasc{JRASC}%
\def\memras{MmRAS}%
\def\mnras{MNRAS}%
\def\pra{Phys.~Rev.~A}%
\def\prb{Phys.~Rev.~B}%
\def\prc{Phys.~Rev.~C}%
\def\prd{Phys.~Rev.~D}%
\def\pre{Phys.~Rev.~E}%
\def\prl{Phys.~Rev.~Lett.}%
\def\pasp{PASP}%
\def\pasj{PASJ}%
\def\qjras{QJRAS}%
\def\skytel{S\&T}%
\def\solphys{Sol.~Phys.}%

\def\sovast{Soviet~Ast.}%
\def\ssr{Space~Sci.~Rev.}%
\def\zap{ZAp}%
\def\nat{Nature}%
\def\iaucirc{IAU~Circ.}%
\def\aplett{Astrophys.~Lett.}%
\def\apspr{Astrophys.~Space~Phys.~Res.}%
\def\bain{Bull.~Astron.~Inst.~Netherlands}%
\def\fcp{Fund.~Cosmic~Phys.}%
\def\gca{Geochim.~Cosmochim.~Acta}%
\def\grl{Geophys.~Res.~Lett.}%
\def\jcp{J.~Chem.~Phys.}%
\def\jgr{J.~Geophys.~Res.}%
\def\jqsrt{J.~Quant.~Spec.~Radiat.~Transf.}%
\def\memsai{Mem.~Soc.~Astron.~Italiana}%
\def\nphysa{Nucl.~Phys.~A}%
\def\physrep{Phys.~Rep.}%
\def\physscr{Phys.~Scr}%
\def\planss{Planet.~Space~Sci.}%
\def\procspie{Proc.~SPIE}%
\def\pasa{PASA}
%


\begin{abstract}
 {
 Black holes with masses of  $\rm 10^6-10^9~M_{\odot}$ dwell in the centers of most galaxies, but their formation mechanisms are not well known. A subdominant dissipative component of  dark matter with similar properties to the ordinary baryons, known as mirror dark matter, may collapse to form massive black holes during the epoch of first galaxies formation. In this study, we explore the possibility of massive black hole formation via this alternative scenario.  We  perform three-dimensional  cosmological simulations  for  four distinct halos and  compare their thermal, chemical and dynamical evolution in both the ordinary and the mirror sectors.  We find that the collapse of halos is significantly delayed  in the mirror sector  due to the lack of $\rm H_2$ cooling and only halos with masses above $ \rm \geq 10^7~ M_{\odot}$ are formed.  Overall, the mass inflow rates  are $\rm \geq 10^{-2}~M_{\odot}/yr$ and there is  less fragmentation. This suggests that the conditions for the formation of massive objects, including black holes, are more favorable in the mirror sector.
 
  } 
 \end{abstract}



\section{Introduction}
Most of the galaxies if not all  today harbor supermassive black holes (SMBHs) of a few million to billion solar masses \citep{Kormendy2013} and their presence has also been revealed from the observations of quasars up to $z \geq 7$, a few hundred million years after the Big Bang \citep{Fan2003,Willott2007,Jiang2009,MOrtlock2011,Venemans2015,Wu2015,Banados18,Schleicher18}. The existence of such massive objects at  early epochs poses a challenge to our understanding of structure formation in the Universe.  How come they formed and how did they grow  are still open questions.

Various models of black hole (BH) formation have been proposed in the literature, which include the collapse of stellar remnants, runaway collisions in stellar clusters and  the collapse of a giant gas cloud into a massive black hole, i.e. the so-called direct collapse model. These models provide seed BHs of $\rm 10-10^5~M_{\odot}$ and  these seed BHs have to efficiently grow to reach the observed masses within the first billion years.  Population III stars, depending upon their mass, may collapse into a BH of a few hundred solar masses. However, they have to  continuously grow at the Eddington limit  to reach the observed masses. The feedback from BHs halts the accretion onto them \citep{Johnson2007, Alvarez2009,Smith17} and they  may require a few episodes of super-Eddington accretion \citep{Madau14,Mayer2015,Lupi16,Inayoshi16} to grow to a billion solar masses.  The mass of BHs resulting from the runaway collisions in dense stellar clusters depends  on the density, metallicity and the initial mass of the cluster. Under optimal conditions, seed BHs from this scenario can have masses of about a thousand solar masses and have to form within the first 2-3 million years  \citep{Zwart2002,Devecchi2012,Katz2015,Reinoso18a,Sakurai18,Reinoso18b}. Particularly,  the potential interaction between stellar collisions and gas accretion is important. The latter may enhance the black hole mass formed in the first stellar clusters \citep{Boekholt18}, or trigger the formation of run-away mergers in clusters of stellar mass black holes \citep{Davies11,Lupi14}. The direct collapse model, on the other hand, provides BH seeds with masses of about $\rm 10^5~ M_{\odot}$, but requires  large inflow rates of about $\rm 0.1~M_{\odot}/yr$ \citep{Schleicher13,LatifViscous2015,Latif2016}.  Such conditions can be achieved  in metal free halos illuminated by strong UV radiation \citep{Chon17}. However, it is still not clear what would be the number density of direct collapse black holes. Even the massive seeds would require rather special conditions to efficiently grow, see \cite{Latif18}  and \cite{Regan18}. Each model has its pros and cons, further details about the models can be found in  the dedicated reviews on this topic \citep{Volonteri2012,Haiman2013,Latif16PASA,Woods18}. 


In this article, we take an alternative approach and explore the possibility of forming massive BHs via dissipative Dark Matter (DM). \cite{Damico18} (hereafter called D18)  have proposed that a small component of mirror matter (i.e.~an elegant model of dissipative DM; see e.g.~\citep{Berezhiani05,Foot:2004pa,Blinnikov:1983gh,Khlopov:1989fj}) may form intermediate mass BHs. They have shown, using one-zone models, that the thermal and the chemical evolution for both the mirror $\mathcal{M}$ and the ordinary $\mathcal{O}$ sectors are different due to the lower $\mathcal{M}$ radiation temperature. Furthermore, they found that the thermal evolution in the $\mathcal{M}$  sector depends on the virial temperature of the halo and 3D simulations are required to investigate this effect. They also pointed out that, in the presence of this dissipative DM sector, the BHs are expected to grow at a faster rate with respect to the ordinary case as they can accrete  both collapsed $\mathcal{O}$  and $\mathcal{M}$  matters.  Motivated by the work of D18, we perform 3D cosmological simulations of the first minihalos forming at $z=20-30$ and explore the impact of hydrodynamics and the collapse dynamics on the thermal and chemical properties of the halos in the $\mathcal{M}$  sector. We compare our results with the $\mathcal{O}$  sector and also assess the inflow rates, as well as study fragmentation properties of these halos. Our findings suggest that halos forming in the $\mathcal{M}$  sector are about an order of magnitude more massive and they may have important implications for the formation of first structures in the Universe.

Our article is organized as follows. In section two, we describe the model, and the numerical methods and initial conditions.  We present our results in section three and confer our conclusions in section four.

\section{Setting the stage}
\subsection{A dissipative DM Model}
In terms of microphysical properties, a symmetric mirror sector (the parity symmetry which exchanges $\mathcal{M}$ and $\mathcal{O}$ field is not broken) is identical to the standard model of particle physics, for details see D18 and references therein.  The $\mathcal{M}$ sector differs from the $\mathcal{O}$ one only in two macroscopic quantities: $i)$ the ratio of abundances $\beta = \Omega^{'}_{\rm b} /\Omega_{\rm b}$ and $ii)$ the ratio of radiation temperatures $x = T^{'}_{ \gamma} /T_{ \gamma}$. The $'$ symbol denotes mirror matter. To avoid stringent BBN and CMB limits, the $\mathcal{M}$ sector must be colder than the ordinary one ($x\lesssim 0.3$~\citep{Berezhiani:2000gw,Berezhiani05,Foot:2014uba}). 

In our simulations, we assume $\beta = 1$ and $x=0.01$. The rest of DM is in the form of standard cold and collisionless DM. These values could result from a $\mathcal{M}$ sector with a broken mirror parity (see e.g.~\cite{Berezhiani:1995am}) or minimal mirror twin Higgs (see e.g.~\cite{Barbieri:2016zxn}).  Although one cannot easily predict typical values for these parameters, we expect that our benchmark model exhibits macrophysical behavior common to several microphysical realizations of mirror DM. In addition to the $\beta$ and $x$ parameters, we assume for simplicity that the chemistry is the same in the two sectors. However, one can easily imagine scenarios in which the mirror chemistry makes production of  $\rm H_2$ more difficult \citep{Rosenberg17}. We cannot study these at the moment, but we expect that our benchmark model exhibits similar qualitative features.

We neglect the baryonic component when adding the $\mathcal{M}$  sector, effectively including only conventional DM and mirror DM in the simulations. These components only interact gravitationally (and via renormalizable portals that we set to zero), so we assume their impact is not strong as on larger scales gravity is dominated by the DM, and on smaller scales self-gravity takes over. To solve the chemical and thermal evolution of  $\mathcal{M}$  matter in the early Universe, we follow the approach of D18. They have shown that due to the faster recombination in the $\mathcal{M}$  sector, the fraction of free electrons is lower and results in a suppressed abundance of $\mathcal{M}$ molecular hydrogen. 
\begin{figure*}
\begin{tabular}{c}
\begin{minipage}{7cm}
\vspace{-0.0cm}
\hspace{-5cm}
\includegraphics[scale=0.8]{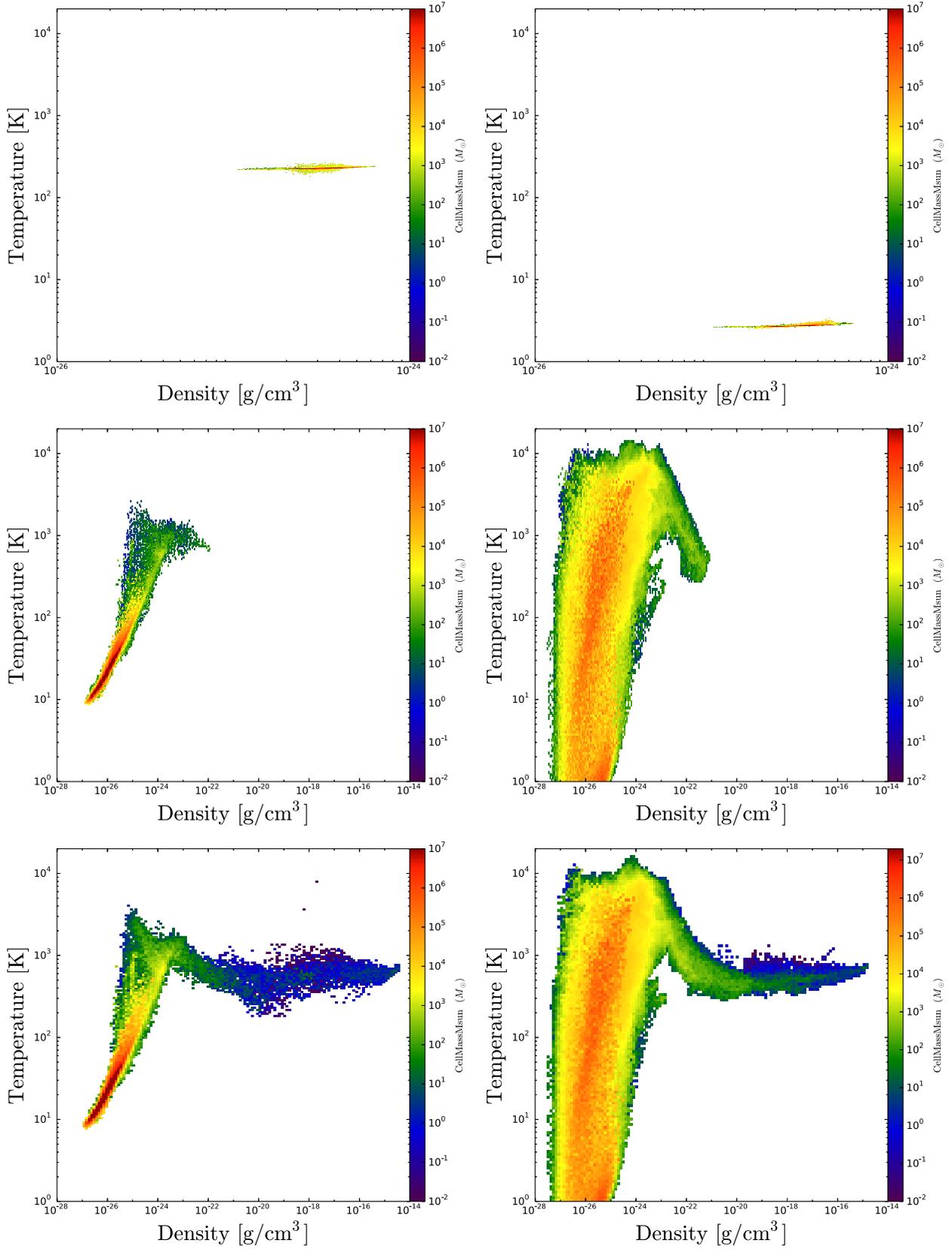}
\end{minipage} 
\end{tabular}
\caption{Phase plots of  the  temperature in the  reference halo for both the $\mathcal{O}$ (left) and the $\mathcal{M}$ (right) sectors. The top row shows the initial redshift, the middle row the virialization redshift  and the bottom row the collapse redshift.}
\label{fig1}
\end{figure*}

\subsection{Numerical methods}
We employ the publicly available code Enzo~\citep{Enzo2014} to conduct hydrodynamical cosmological simulations. Enzo is an open source, adaptive mesh refinement (AMR), parallel, multi-physics simulation code which can run and scale well on various platforms using the message passing interface (MPI). Hydrodynamics is solved using the piece-wise parabolic method (PPM) and the DM dynamics is computed with particle-mesh based N-body solver. For self-gravity calculations, we use a multi-grid Poisson solver.

We make use of the MUSIC package~\citep{Hahn2011} to generate  cosmological initial conditions typically at $z \geq 100$  by using the PLANCK 2016 data with $\Omega_{\rm M}=0.3089$, $\Omega_{\Lambda}=0.6911$, $\rm H_{0}=0.6774$ \citep{Planck2016}. Our cosmological volume (simulation box) has a comoving size of 1 Mpc/h, we select the most massive halo forming in our computational domain at $z \ge 20$ and place it at the center of the box. We employ nested grid initial conditions with top-level grid resolution of $\rm128^3$ cells and an equal number of  DM particles. We subsequently employ two additional nested grids each with the same resolution as of a top grid. In addition to this, we further employ 18 additional levels of refinement during the course of the simulations which provide us with an effective resolution of about 200 AU. We ensure a Jeans resolution of at least four cells during the simulations. In total, we employ 5767168 DM particles to solve the N-body dynamics which provides us an effective DM resolution of about a few hundred solar masses. Our refinement criterion is based on the baryonic overdensity and the DM mass resolution. A cell is marked for refinement when it exceeds four times the cosmic mean density or DM particle density of 0.0625 times $\rho_{DM}r^{\ell \alpha}$  where $\rho_{DM}$ is the dark matter density, $r = 2$ is the refinement factor, $\ell$ is the refinement level, and $\alpha = -0.3$ makes the refinement super-Lagrangian. 

We use the KROME package~\citep{Grassi2014} to self-consistently solve the thermal and chemical evolution of nine primordial species {\bf (}$\rm H,~H^+,~ H^-, ~He,~ He^+, ~He^{++},~ H_2, ~H_2^+, ~e^-${\bf )}\footnote{From now on, instead of using chemical notation for $\rm H^+$ and $\rm H$, we use HII and HI, respectively.} in cosmological simulations.  Our chemical model includes the most important gas phase reactions and processes including the formation of molecular hydrogen. It also includes the cooling and heating processes due to collisional excitation, collisional ionization, radiative recombination, collisional induced emission,  $\rm H_2 $ and  chemical heating/cooling.

For the $\mathcal{M}$  sector with $x=0.01$, the $\mathcal{M}$ helium fraction is almost negligible~\citep{Berezhiani05}. More importantly,  the initial abundance of free $\mathcal{M}$  electrons at $z=100$ is about four orders of magnitude smaller than the ordinary one (see Fig. 1 of D18) and we scale the fractions of other species accordingly.

\begin{table*}
\begin{center}
\caption{The properties of the simulated halos are listed here. The second and fourth columns are the halo masses w/o (Halo Mass$^{\mathcal{O}}$) and w/ (Halo Mass$^{\mathcal{M}}$) mirror baryons. The third and fifth columns denote the collapse redshifts in the $\mathcal{O}$ and $\mathcal{M}$ sectors respectively.}
\begin{tabular}{| c | c | c | c |c |}
\hline
\hline
 
Model   & Halo Mass$^{\mathcal{O}}$  & Collapse redshift$^{\mathcal{O}}$ &  Halo Mass$^ {\mathcal{M}}$   & Collapse redshift$^{\mathcal{O}}$ \\

No &   $\rm M_{\odot} $ & $z$  &  $\rm M_{\odot}$ & $z$   \\
\hline                                                          \\

1     & $\rm 3.5 \times 10^{5}$ & 23.3 & $\rm 1.01 \times 10^{7}$  & 13.79\\
2     & $\rm 1.3 \times 10^{6}$ & 20  & $\rm 2.69 \times 10^{7}$  & 14.05\\
3     & $\rm 7.8 \times 10^{5}$ & 24.5 & $\rm 1.3 \times 10^{7}$  & 13.35\\
4     & $\rm 6.7  \times 10^{5}$ & 25.4 & $\rm 1.2 \times 10^{7}$  & 16\\
\hline
\end{tabular}
\label{table1}
\end{center}
\end{table*}

\begin{figure*}
\begin{tabular}{c}
\begin{minipage}{8cm}
\vspace{0.0cm}
\hspace{-5cm}
\includegraphics[scale=0.84,trim={0cm 1.0cm 0cm 1cm},clip]{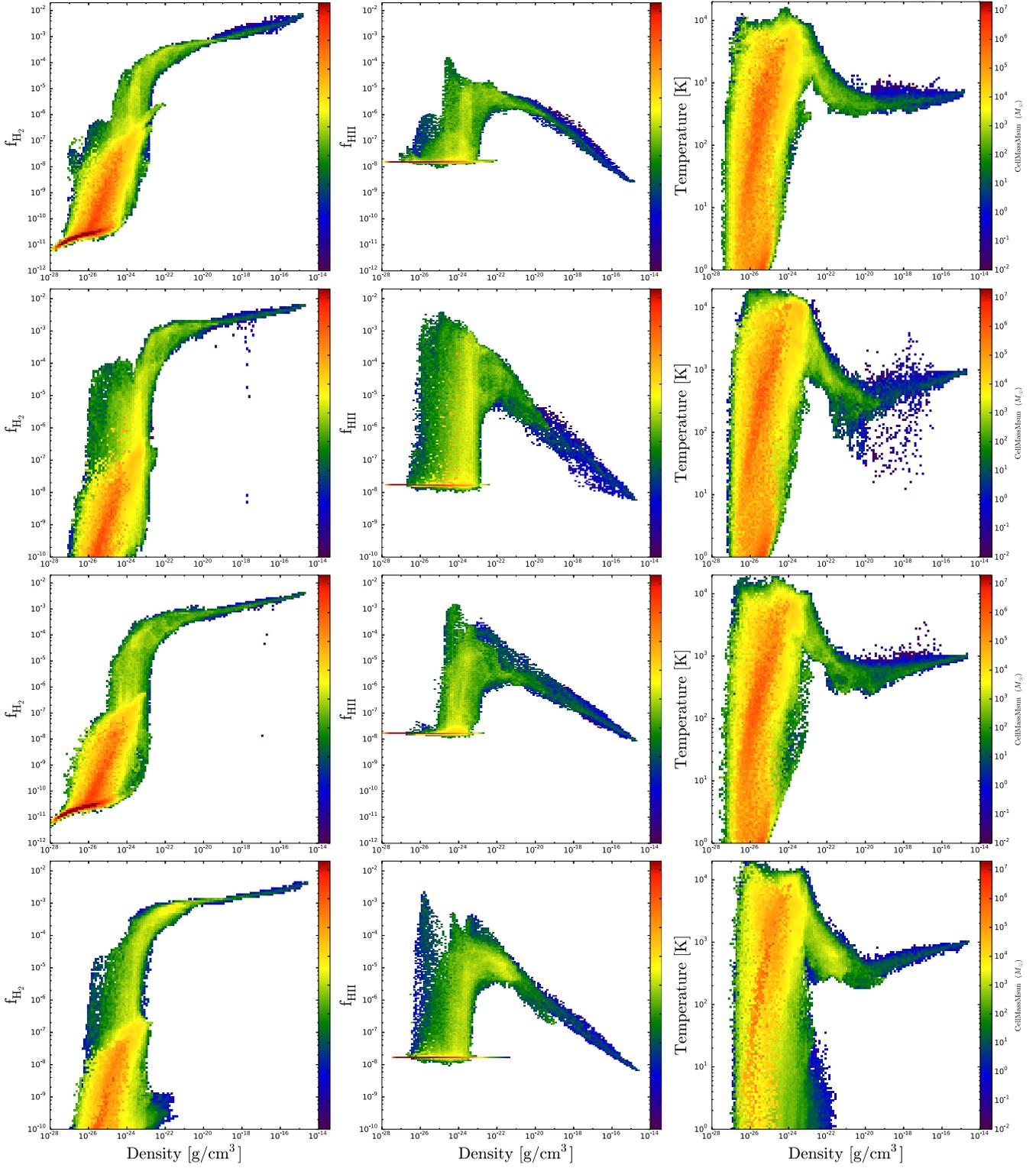}
\end{minipage} 
\end{tabular}
\caption{Phase plot of $\rm H_{2}$  and $\rm HII$ mass fractions, and  temperature against  gas density for four halos in the mirror sector.  Each row represents a different halo at its collapse redshift, as listed in table 1.}
\label{fig2}
\end{figure*}

 \begin{figure*} 
\begin{tabular}{c c}
\begin{minipage}{6cm}
\hspace{-1.5cm}
\includegraphics[scale=0.4]{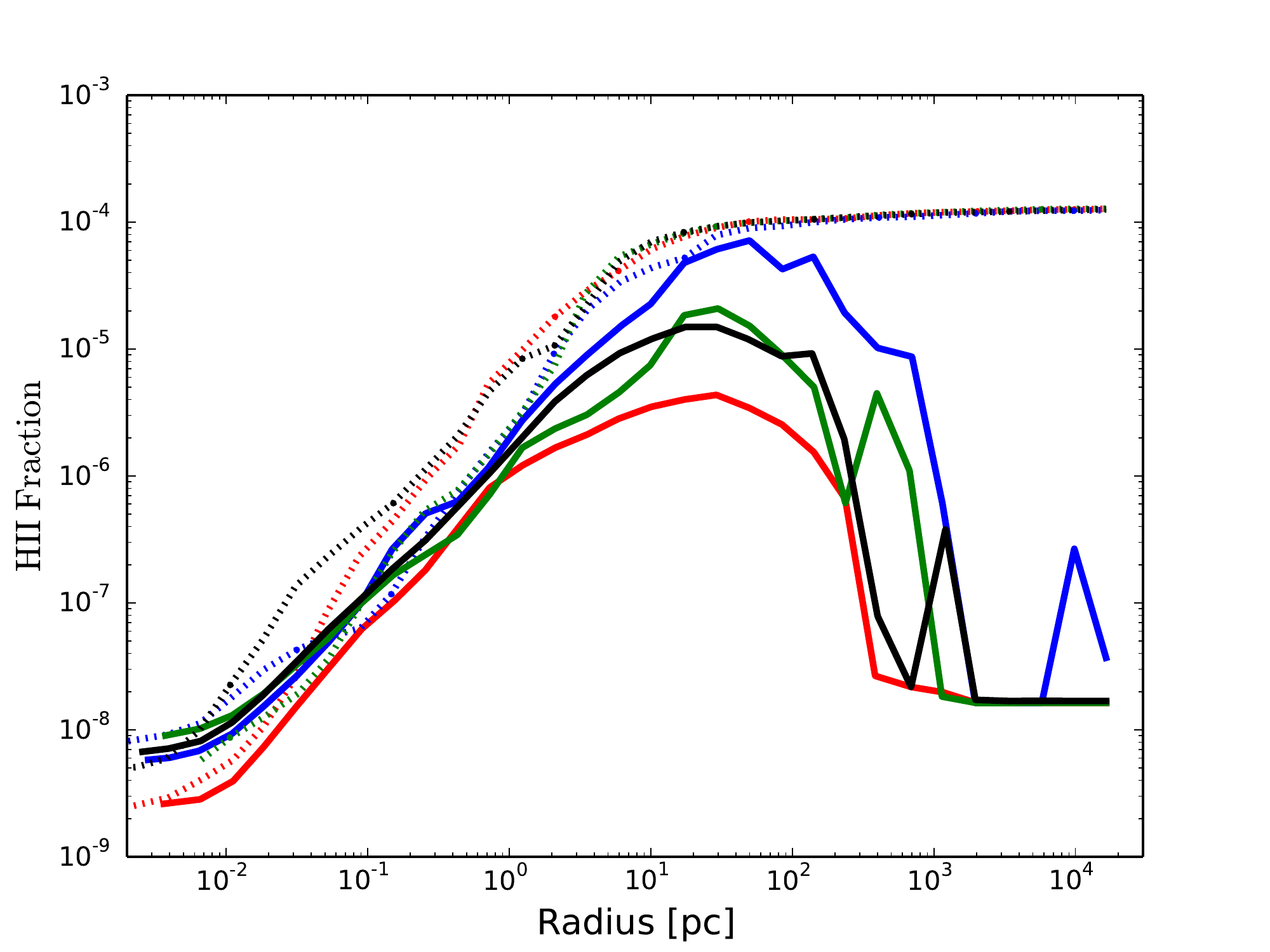}
 \end{minipage} &
 \begin{minipage}{6cm}
\hspace{-0.0cm}
\includegraphics[scale=0.4]{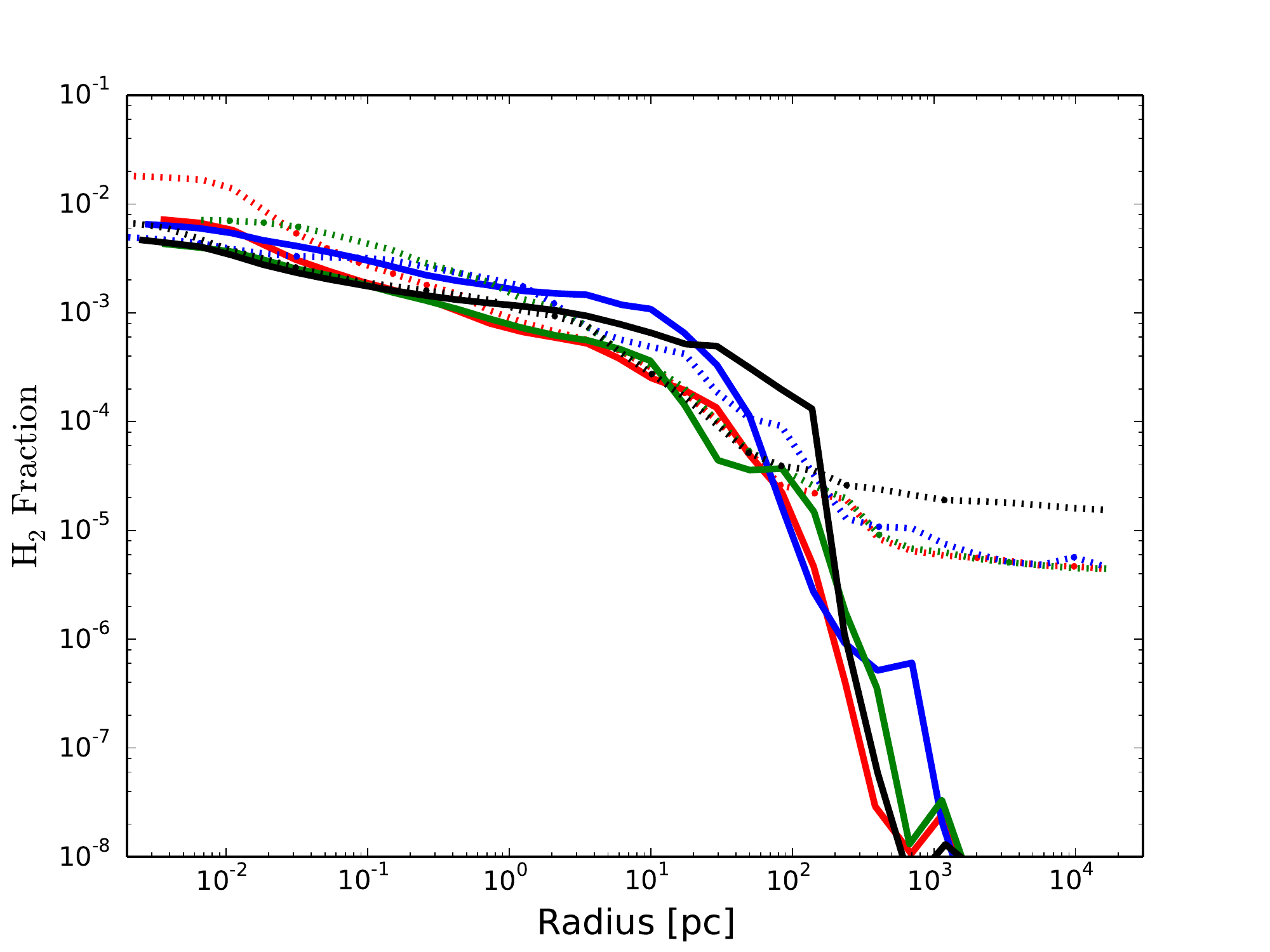} 
\end{minipage} 
\end{tabular}
\caption{Spherically averaged and radially binned profiles of the $\rm HII$, and $\rm H_2$ mass fractions.  The solid lines represent halos in the mirror sector while dashed lines are for the halos in the ordinary sector, as listed in table 1, all computed at the collapsed redshift. The colors, red, blue, green and black represent halos 1, 2, 3 and 4, respectively.}
\label{fig3}
\end{figure*}

\begin{figure*} 
\begin{center}
\includegraphics[scale=0.6]{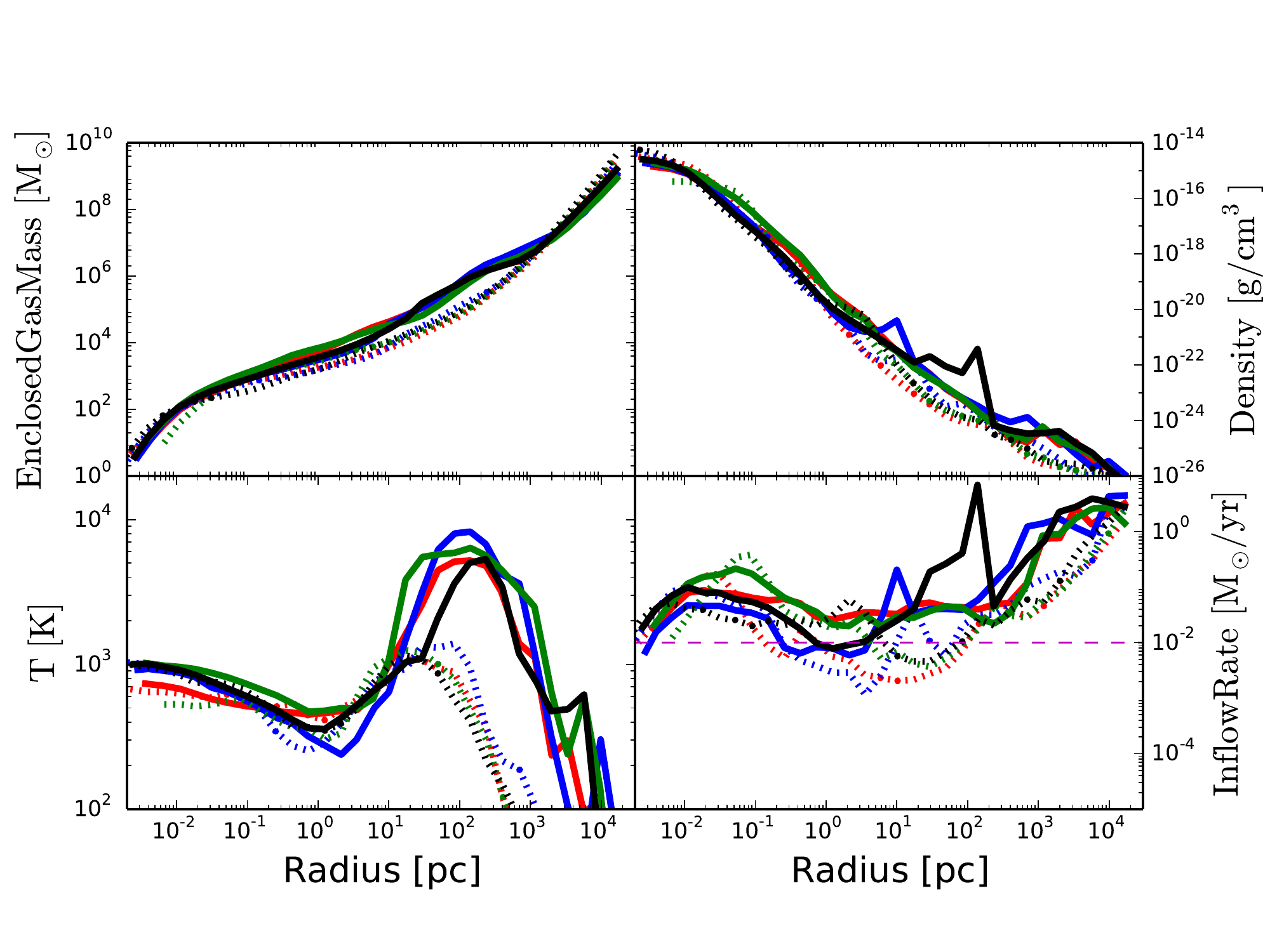}
\end{center}
\caption{Spherically averaged and radially binned profiles of gas density, temperature, enclosed gas mass  and mass inflow rates.  The solid lines represent halos in the mirror sector while dashed lines are for the halos in the ordinary sector, all computed at the collapsed redshift, as listed in table 1. The colors, red, blue, green and black represent halos 1, 2, 3 and 4, respectively.}
\label{fig4}
\end{figure*}

\section{Results}
We present the main findings of the present work in this section. In total, we have performed eight cosmological simulations for four different halos in both sectors. The properties of the simulated halos are listed in table \ref{table1}. In the coming  subsections, we  discuss  and compare the  thermal  and chemical evolution  of the halos in both sectors.  We also  point out the  differences in the fragmentation properties of  the halos. 

\subsection {Time evolution of a reference run in the mirror and baryonic sector}
We take halo 1 as the reference case and compare its thermal evolution in both sectors as shown in figure \ref{fig1}.  The temperature of the gas in the $\mathcal{M}$ sector at $z=100$ is lower than the $\mathcal{O}$ sector due to  inefficient Compton heating. As collapse proceeds gas falls in the DM potential and  gets heated up to  about 1000 K  due to the virialization shocks.  In the $\mathcal{O}$ sector, gas starts to cool and collapse  after reaching the  molecular hydrogen cooling  threshold (a few times $\rm 10^5~M_{\odot}$)  at $z$ =23.  In the $\mathcal{M}$ sector, due to the inefficient production of molecular hydrogen, gas cannot cool until the halo mass reaches the atomic cooling limit.  Consequently, the halo virial temperature reaches around $\rm 10^4~K$ and the halo mass  $\rm \geq 10^7~M_{\odot}$.  For the $\mathcal{M}$ sector, strong shocks during the virialization of an atomic cooling halo catalyze  $\rm H_2$ formation by enhancing the electron fraction and as a result $\rm H_2$ abundance gets significantly increased. 

Once sufficient $\mathcal M$ H$_2$  has formed,  the gas temperature  is brought down to about a few hundred Kelvin in the core of  the atomic cooling halo. Overall, the temperature in the center of $\mathcal{M}$ halos  is about a factor of two higher compared to the $\mathcal{O}$ sector.  The halo mass  is  about a  factor of twenty  larger  and the collapse is delayed until $ z \sim 14 $, see table \ref{table1}.  This suggests that the first halos forming in the $\mathcal{M}$ sector are atomic cooling halos while in the $\mathcal{O}$ sector minihalos are formed first. This will have an important implications for the early structure formation, see our discussion below.


\subsection {Comparison of different halos and fragmentation study}
We here compare the chemical  and thermal evolution of four different halos in the $\mathcal{M}$ sector only.  The  plots of $\rm H_2$ and  $\rm HII$ mass fractions  and  temperature  against the gas density  at the collapse redshifts of the halos are shown in figure \ref{fig2}.  The HII fraction at low densities is about  $\rm 10^{-7}$, a few orders of magnitude lower compared to the $\mathcal{O}$ sector and reaches up to $\rm \sim 10^{-4}$   at densities between $\rm 10^{-26}-10^{-24}~g/cm^3$ due to the strong virialization shocks.  At higher densities, the $\rm HII$ fraction starts to decline due to the Lyman alpha and molecular hydrogen cooling which brings the  temperature down. This trend has been observed for all halos. The enhanced electron fraction during the process of virialization  acts as a catalyst for the formation of molecular hydrogen and as a result the $\rm H_2$ fraction gets boosted  about six orders of magnitude.  After the halo has virialized, the $\rm H_2$ fraction continues to increase and  gets further enhanced due to the three body reactions. The typical abundance of $\rm H_2$ in the core of the halo is a few times $\rm 10^{-2}$ and the same is for all halos. Overall, the halos with higher virial masses have  higher HII and $\rm H_2$ fractions.
 
Contrary to the $\mathcal{O}$ sector, the temperature of  the halo continues  to increase until it reaches the  atomic cooling regime due to the low $\rm H_2$ fraction at earlier times.  By that  time, the  halo mass is above $\rm 10^7~M_{\odot}$ and Lyman alpha cooling becomes effective at densities above $\rm 10^{-24}~g/cm^3$.  After  reaching the atomic cooling limit, the molecular hydrogen formed during the virialization  brings the gas temperature down to a few hundred K.  Consequently, the formation of minihalos  remains suppressed in the $\mathcal{M}$ sector.
To further clarify the differences  between two sectors,  we  show the averaged radial profiles of  $\rm H_2$ and $\rm HII$ mass fractions in figure \ref{fig3}.  Although the initial abundances of $\rm H_2$, $\rm HII$, HI and electrons are a few orders of magnitude lower in the $\mathcal{M}$ sector,  they become almost similar to the $\mathcal{O}$ sector after virialization.  Due to the larger halo mass in $\mathcal{M}$ sector, stronger virialization shocks  boost the abundances of these species and  reduce the differences  between the two sectors. 

 To  compare the dynamical properties of the halos,  we show the profiles of the temperature,  density, enclosed mass and the mass inflow rates  in figure \ref{fig4}.  The density in  the outskirts of the halo is  $\rm 10^{-24}~g/cm^3$ and increases up to $\rm 10^{-16}~g/cm^3$ in the core, with small bumps due to the presence of substructure. The density profiles are almost similar in both sectors  for all halos and follow  the $\sim R^{-2.1}$ behaviour as expected for $\rm H_2$ cooled gas. Consequently, the enclosed gas mass profiles are similar  in both sectors, but  almost an order of magnitude larger around 100~pc in the $\mathcal{M}$ sector.  This difference comes from the larger halo masses in the $\mathcal{M}$ sector.  The temperature profiles show that differences between the two sectors are very prominent above 10 pc. For instance, at 100 pc the temperature in the $\mathcal{M}$ sector is about $\rm 8000-10^4$ K while in the $\mathcal{O}$ sector the temperature does not exceed $\rm 10^3$ K. These differences are  again due to the larger halo masses in the $\mathcal{M}$ sector.  In general, gas in the centre of the halos is warmer  in the $\mathcal{M}$ sector. The average mass inflow rates for both sectors are between $\rm 0.001- 0.1~M_{\odot}/yr$ and  they are generally lower in the $\mathcal{O}$ sector. 

\begin{figure*}
\begin{tabular}{c}
\begin{minipage}{7cm}
\hspace{-5cm}
\includegraphics[scale=0.8,trim={0cm 8.0cm 0.5cm 8cm},clip]{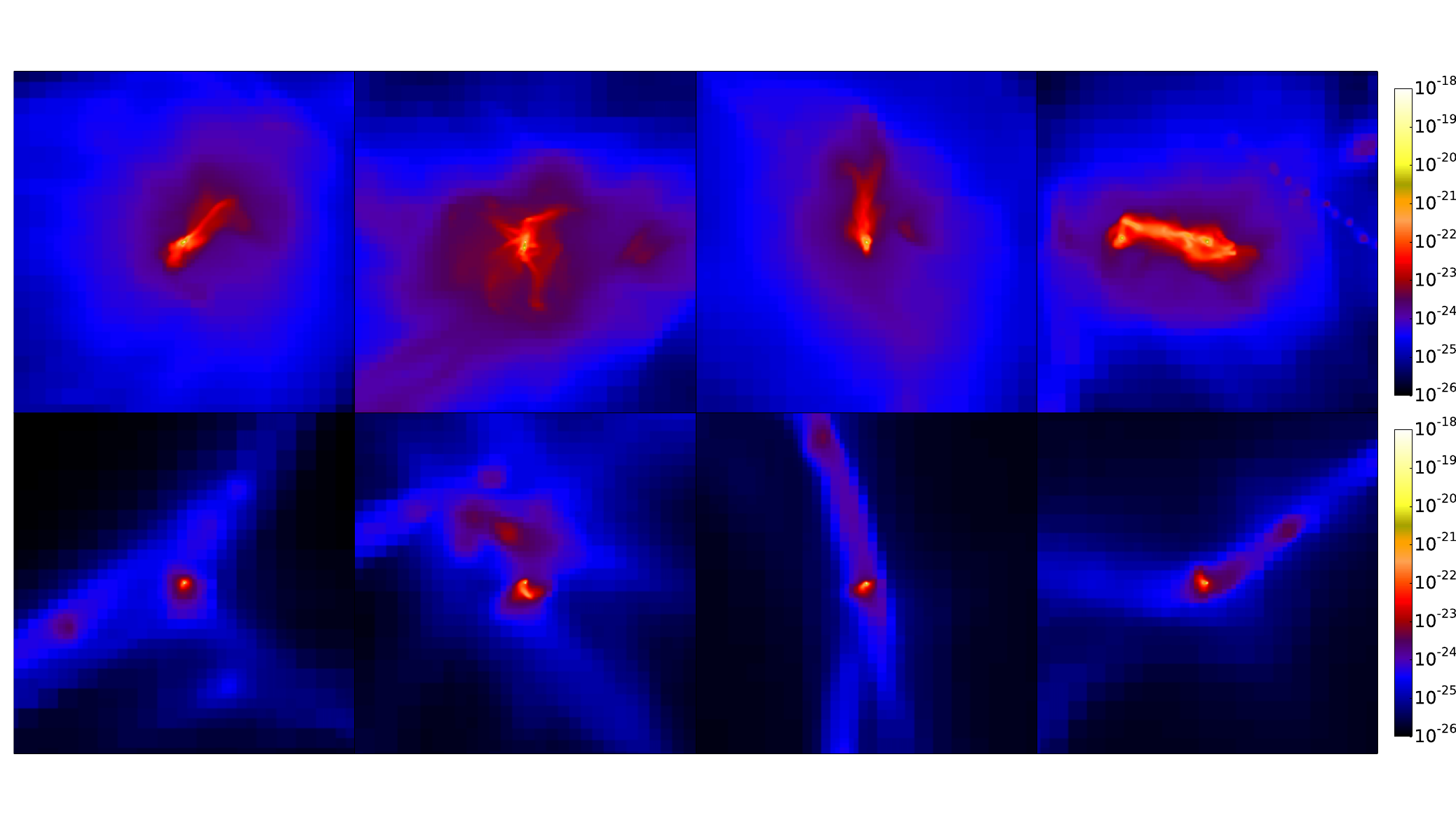} 
\end{minipage} 
\end{tabular}
\caption{Density projections (the average density along the line of the sight) at the collapse redshifts of the halos for the central 500 pc region. The top row shows halos in the mirror sector while the bottom row corresponds to the ordinary sector.}
\label{fig5}
\end{figure*}


The morphology of the  halos  at their collapse redshifts  is shown in figure \ref{fig5}.  The density  structure in the central region of the halo is different for both sectors  and this trend is observed for all halos. In general,  for  the $\mathcal{O}$ sector the density structure is  more dense, filamentary and compact compared to the $\mathcal{M}$ one.  Moreover,  there are more than one gas clumps, and they are well separated, while in the $\mathcal{M}$ sector structures are more spherical and fluffy. This comes from the fact that the molecular hydrogen is  mainly  concentrated in the core of the halo in the $\mathcal{M}$ sector  while in the outskirts of the halo $\rm H_2$ is below the universal value (i.e. $\rm 10^{-3}$). Based on these indications,  more fragmentation is expected  in the $\mathcal{O}$ sector.  However, the possibility of fragmentation at the later stages of collapse in the $\mathcal{M}$ sector cannot be completely ruled out.

\section{Discussion and conclusions}

In this study, we have investigated the possibility of BH formation in the context of dissipative DM. Previous works (D18) suggest that a small component of  DM  similar to the baryonic matter may collapse to form massive BHs. D18 show, employing one-zone models, that the evolution in the $\mathcal{M}$ sector is very different from the $\mathcal{O}$ one. Motivated by the work of D18, we have performed 3D cosmological simulations for four different halos to explore the impact of hydrodynamics and collapse dynamics on the thermal and chemical evolution as well as their implications for structure formation. 

In our simulations of $\mathcal{O}$ sector, we have only ordinary baryons and collisionless DM  to compare with simulations of $\mathcal{M}$ one. For the $\mathcal{O}$ sector, the mirror component is expected to collapse later than the baryonic component and therefore to have negligible effect. For the $\mathcal{M}$ sector, the potential influence of the baryons is only via gravity. Hence we expect that due to the inefficient cooling, the cloud will not be able to collapse even if we take into account the additional gravitational effect of baryons.

\medskip 
In general our results are in an agreement with the findings of D18.  We show that the formation of minihalos remains suppressed in the $\mathcal{M}$  sector due to the deficiency of molecular hydrogen and the gravitational collapse is significantly delayed by $ \Delta z \sim 10$. Consequently, gas keeps collapsing until the halo mass reaches an atomic cooling limit with typical halo masses of $\rm  10^{7} ~M_{\odot}$ and virial temperatures around $\rm 10^4~K$. Before the virialization of the halos, the abundances of $\rm H_2$ and $\rm HII$ are a few orders of magnitude lower in the $\mathcal{M}$  sector, but become comparable to the $\mathcal{O}$  one after virialization because of strong shocks. In general, the $\rm H_2$ mass fraction is about a factor of two lower and the temperature is about a factor of two higher. The mass inflow rate is $\rm \geq 10^{-2}~M_{\odot}/yr$, a factor of a few higher than in the baryonic sector. The degree of fragmentation is also very low. 
Overall, halos in the $\mathcal{M}$ sector are very similar to halos irradiated by a moderate background UV flux in the standard scenario \citep{Schleicher10,Latif2016dust}. 

These factors suggest that the conditions for the formation of massive objects, including BHs, are more favourable in the $\mathcal{M}$ sector. Furthermore, when BHs form the accretion rate is largely boosted because they can accrete a substantial portion of dissipative DM. Our results reinforce the findings of D18 and provide a viable alternative of massive BH formation at high redshift.

%
%

\section*{Acknowledgments}
ML thanks the UAEU for funding via startup grant No..... AL acknowledges support from the European Research Council project No. 740120 'INTERSTELLAR'. DRGS thanks for funding via Conicyt PIA ACT172033, Fondecyt regular (project code 1161247), the ''Concurso Proyectos Internacionales de Investigaci\'on, Convocatoria 2015'' (project code PII20150171) and the BASAL Centro de Astrof\'isica y Tecnolog\'ias Afines (CATA) PFB-06/2007.  GDA is supported by the Simons Foundation Origins of the Universe program (Modern Inflationary Cosmology collaboration). This work has made use of the Horizon Cluster, hosted by Institut d'Astrophysique de Paris, to carry out and analyse the presented simulations. SB is financially supported by Fondecyt Iniciacion (project code 11170268), CONICYT programa de Astronomia Fondo Quimal 2017 QUIMAL170001, and BASAL Centro de Astrofisica y Tecnologias Afines (CATA) AFB-17002.

%
 \bibliography{smbhs.bib}
 
\newpage

\end{document}